\documentstyle[12pt]{article}
\begin{document}

\pagenumbering{arabic}

\def\LAMBDA{\mbox{\rlap{$\raise3pt\hbox{--}$}{$\lambda$}}}
\def\nn{\nonumber}
\def\ba{\begin{array}}
\def\ea{\end{array}}
\def\B{{\mbox{\boldmath $B$}}}
\def\p{{\mbox{\boldmath $p$}}}
\def\hatp{{\widehat{\mbox{\boldmath $p$}}}}
\def\vpi{{\mbox{\boldmath $\pi$}}}
\def\r{{\mbox{\boldmath $r$}}}
\def\Nab{{\mbox{\boldmath $\nabla$}}}
\def\ddz{\frac{\partial}{\partial z}}

\begin{center}

{\LARGE\bf
Analogies between light optics and charged-particle optics
}

\bigskip

{\em Sameen Ahmed KHAN} \\

\bigskip

khan@fis.unam.mx,  ~ rohelakhan@yahoo.com \\
http://www.pd.infn.it/$\sim$khan/ \\
http://www.imsc.ernet.in/$\sim$jagan/khan-cv.html \\
Centro de Ciencias F\'{i}sicas, \\
Universidad Nacional Aut\'onoma de M\'exico (UNAM) \\
Apartado Postal 48-3,
Cuernavaca 62251,
Morelos, \\
{\bf M\'EXICO} \\

\end{center}

\medskip
\medskip

\begin{abstract}
The close analogy between geometrical optics and the classical theories
of charged-particle beam optics have been known for a very long time.
In recent years, quantum theories of charged-particle beam optics have
been presented with the very expected feature of wavelength-dependent
effects.  With the current development of non-traditional prescriptions
of Helmholtz and Maxwell optics respectively, accompanied with the
wavelength-dependent effects, it is seen that the analogy between the
two systems persists.  A brief account of the various prescriptions
and the parallel of the analogies is presented.
\end{abstract}

\def\i{{\rm i}}

\section{Introduction}
Historically, variational principles have played a fundamental role
in the evolution of mathematical models in classical physics, and many
equations can be derived by using them.  Here the relevant examples are
Fermat's principle in optics and Maupertuis' principle in mechanics.
The beginning of the analogy between geometrical optics and mechanics
is usually attributed to Descartes (1637), but actually it can traced
back to Ibn Al-Haitham Alhazen (0965-1037)~\cite{Ambrosini}.  The
analogy between the trajectory of material particles in potential
fields and the path of light rays in media with continuously variable
refractive index was formalized by Hamilton in 1833.  This Hamiltonian
analogy lead to the development of electron optics in 1920s, when Busch
derived the focusing action and a lens-like action of the axially
symmetric magnetic field using the methodology of geometrical optics.
Around the same time Louis de Broglie associated his now famous
wavelength to moving particles.  Schr\"{o}dinger extended the analogy
by passing from geometrical optics to wave optics through his wave
equation incorporating the de Broglie wavelength.  This analogy played
a fundamental role in the early development of quantum mechanics.  The
analogy, on the other hand, lead to the development of practical
electron optics and one of the early inventions was the electron
microscope by Ernst Ruska.  A detailed account of Hamilton's analogy
is available in~\cite{Hawkes}-\cite{Forbes}.

Until very recently, it was possible to see this analogy only between
the geometrical-optic and classical prescriptions of electron optics.
The reasons being that, the quantum theories of charged-particle beam
optics have been under development only for about a
decade~\cite{JSSM}-\cite{JK} with the very expected feature of
wavelength-dependent effects, which have no analogue in the traditional
descriptions of light beam optics.  With the  current development of
the non-traditional prescriptions of Helmholtz
optics~\cite{KJS-1,Khan-H-1} and the matrix formulation of Maxwell
optics~\cite{Khan-M-1}-\cite{Khan-M-Review}, accompanied with
wavelength-dependent effects, it is seen that the analogy between the
two systems persists.  The non-traditional prescription of Helmholtz
optics is in close analogy with the quantum theory of charged-particle
beam optics based on the Klein-Gordon equation.  The matrix formulation
of Maxwell optics is in close analogy with the quantum theory of
charged-particle beam optics based on the Dirac equation.  This analogy
is summarized in the table of Hamiltonians.  In this short note it is
difficult to present the derivation of the various Hamiltonians which
are available in the references.  We shall briefly consider an outline
of the quantum prescriptions and the non-traditional prescriptions
respectively.  A complete coverage to the new field of  {\em Quantum
Aspects of Beam Physics} ({\bf QABP}), can be found in the proceedings
of the series of meetings under the same name~\cite{QABP}.

\section{Quantum Formalism}
The classical treatment of charged-particle beam optics has been
extremely successful in the designing and working of numerous optical
devices, from electron microscopes to very large particle accelerators.
It is natural, however to look for a prescription based on the quantum
theory, since any physical system is quantum mechanical at the
fundamental level!  Such a
prescription is sure to explain the grand success of the classical
theories and may also help get a deeper understanding and to lead to
better designing of charged-particle beam devices.

The starting point to obtain a quantum prescription of charged particle
beam optics is to build a theory based on the basic equations  of
quantum mechanics (Schr\"{o}dinger, Klein-Gordon, Dirac) appropriate
to the situation under study.  In order to analyze the evolution of the
beam parameters of the various individual beam optical elements
(quadrupoles, bending magnets,~$\cdots$) along the optic axis of the
system, the first step is to start with the basic time-dependent
equations of quantum mechanics and then obtain an equation of the form
\begin{equation}
\i \hbar \frac{\partial }{\partial s} \psi \left(x , y ;\, s \right)
=
\widehat{\cal H} \left(x , y ;\, s \right)
\psi \left(x , y ;\, s \right)\,,
\label{BOE}
\end{equation}
where $(x , y ;\, s)$ constitute a curvilinear coordinate
system, adapted to the geometry of the system.  Eq.~(\ref{BOE}) is
the basic equation in the quantum formalism, called as the
{\em beam-optical equation}; ${\cal H}$ and $\psi$ as the
{\em beam-optical Hamiltonian} and the {\em beam wavefunction}
respectively.  The second step requires obtaining a relationship
between any relevant observable $\{\langle O \rangle (s) \}$ at the
transverse-plane at $s$ and the observable
$\{\langle O \rangle (s_{\rm in}) \}$
at the transverse plane at $s _{\rm in}$, where $s _{\rm in}$ is some
input reference point.  This is achieved by the integration of the
beam-optical equation in~(\ref{BOE})
\begin{eqnarray}
\psi \left(x , y ; s \right) & = &
\widehat{U} \left(s , s_{\rm in} \right)
\psi \left(x , y ; s_{\rm in} \right)\,,
\label{BOI}
\end{eqnarray}
which gives the required transfer maps
\begin{eqnarray}
\left\langle O \right\rangle \left(s_{\rm in} \right)
\longrightarrow
\left\langle O \right\rangle \left(s \right)
& = &
\left\langle \psi \left(x , y ; s \right)
\left| O \right|
\psi \left(x , y ; s \right) \right\rangle\,, \nn \\
& = &
\left\langle \psi \left(x , y ; s_{\rm in} \right)
\left| \widehat{U} ^{\dagger} O \widehat{U} \right|
\psi \left(x , y ; s_{\rm in} \right) \right\rangle\,.
\label{BOM}
\end{eqnarray}

The two-step algorithm stated above gives an over-simplified picture of
the quantum formalism.  There are several crucial points to be noted.
The first step in the algorithm of obtaining the beam-optical equation
is not to be treated as a mere transformation which eliminates $t$ in
preference to a variable $s$ along the optic axis.  A clever set of
transforms are required which not only eliminate the variable $t$ in
preference to $s$ but also give us the $s$-dependent equation which has
a close physical and mathematical analogy with the original
$t$-dependent equation of standard time-dependent quantum mechanics.
The imposition of this stringent requirement on the construction of the
beam-optical equation ensures the execution of the second-step of the
algorithm.  The beam-optical equation is such that all the required
rich machinery of quantum mechanics becomes applicable to the
computation of the transfer maps that characterize the optical system.
This describes the essential scheme of obtaining the quantum formalism.
The rest is mostly mathematical detail which is inbuilt in the powerful
algebraic machinery of the algorithm, accompanied with some reasonable
assumptions and approximations dictated by the physical considerations.
The nature of these approximations can be best summarized in the optical
terminology as a systematic procedure of expanding the beam optical
Hamiltonian in a power series of $|{\widehat{\vpi}_\perp}/{p_0}|$,
where $p_0$ is the design (or average) momentum of beam particles
moving
predominantly along the direction of the optic axis and
$\widehat{\vpi}_\perp$ is the small transverse kinetic momentum.  The
leading order approximation along with
$|{\widehat{\vpi}_\perp}/{p_0}| \ll 1$, constitutes the paraxial or
ideal behaviour and higher order terms in the expansion give rise to the nonlinear or aberrating behaviour.
It is seen that the paraxial and aberrating behaviour get modified by
the quantum contributions which are in powers of the de Broglie
wavelength ($\bar{\lambda}_0 = {\hbar}/{p_0}$).  The classical limit
of the quantum formalism reproduces the well known Lie algebraic
formalism of charged-particle beam optics~\cite{Lie}.

\section{Light Optics: Non-Traditional Prescriptions}
The traditional scalar wave theory of optics (including aberrations to
all orders) is based on the beam-optical Hamiltonian derived by using
Fermat's principle.  This approach is purely geometrical and works
adequately in the scalar regime.  The other approach is based on the
{\em square-root} of the Helmholtz operator, which is derived from the
Maxwell equations~\cite{Lie}.  This approach works to all orders and
the resulting expansion is no different from the one obtained using
the geometrical approach of Fermat's principle.  As for the
polarization: a systematic procedure for the passage from scalar to
vector wave optics to handle paraxial beam propagation problems,
completely taking into account the way in which the Maxwell equations
couple the spatial variation and polarization of light waves, has been
formulated by analyzing the basic Poincar\'{e} invariance of the system,
and this procedure has been successfully used to clarify several issues
in Maxwell optics~\cite{MSS-1}-\cite{SSM-2}.

In the above approaches, the beam-optics and the polarization are
studied separately, using very different machineries.  The derivation
of the Helmholtz equation from the Maxwell equations is an
approximation as one neglects the spatial and temporal derivatives of
the permittivity and permeability of the medium.  Any prescription
based on the Helmholtz equation is bound to be an approximation,
irrespective of how good it may be in certain situations.  It is very
natural to look for a prescription based fully on the Maxwell
equations, which is sure to provide a deeper
understanding of beam-optics and light polarization in a unified manner.

The two-step algorithm used in the construction of the quantum theories
of charged-particle beam optics is very much applicable in light optics!
But there are some very significant conceptual differences to be borne
in mind.  When going beyond Fermat's principle the whole of optics
is completely governed by the Maxwell equations, and there are no other
equations, unlike in quantum mechanics, where there are separate
equations for, spin-$1/2$, spin-$1$, $\cdots$.

Maxwell's equations are linear (in time and space derivatives) but
coupled in the fields.  The decoupling leads to the Helmholtz equation
which is quadratic in derivatives.  In the specific context of beam
optics, purely from a calculational point of view, the starting
equations are the Helmholtz equation governing scalar optics and for a
more accurate prescription one uses the full set of Maxwell equations,
leading to vector optics.  In the context of the two-step algorithm,
the Helmholtz equation and the Maxwell equations in a matrix
representation can be treated as the `basic' equations, analogue of
the basic equations of quantum mechanics.  This works perfectly fine
from a calculational point of view in the scheme of the algorithm we
have.

Exploiting the similarity between the Helmholtz wave equation and the
Klein-Gordon equation, the former is linearized using
the Feshbach-Villars procedure used for the linearization of the
Klein-Gordon equation.  Then the Foldy-Wouthuysen iterative
diagonalization technique is applied to obtain a Hamiltonian description
for a system with varying refractive index.  This technique is an
alternative to the conventional method of series expansion of the
radical.  Besides reproducing all the traditional quasiparaxial terms,
this method leads to additional terms, which are dependent on the
wavelength, in the optical Hamiltonian.
This is the non-traditional prescription of scalar optics.

The Maxwell equations are cast into an exact matrix form taking into
account the spatial and temporal variations of the permittivity and
permeability.  The derived  representation using $8 \times 8$ matrices
has a close algebraic analogy with the Dirac equation, enabling the use
of the rich machinery of the Dirac electron theory.  The beam optical
Hamiltonian derived from this representation reproduces the
Hamiltonians obtained in the traditional prescription along with
wavelength-dependent matrix terms, which we have named as the
{\em polarization terms}.  These polarization terms are very similar
to the spin terms in the Dirac electron theory and the spin-precession
terms in the beam-optical version of the Thomas-BMT
equation~\cite{CJKP-1}.  The  matrix formulation provides a unified
treatment of beam optics and light polarization.  Some well known
results of light polarization are obtained as a paraxial limit of the
matrix formulation~\cite{MSS-1}-\cite{SSM-2}.
The traditional beam optics is completely obtained from our approach
in the limit of small wavelength, $\LAMBDA \longrightarrow 0$, which
we call as the traditional limit of our formalisms.  This is analogous
to the classical limit obtained by taking $\hbar \longrightarrow 0$,
in the quantum prescriptions.

From the Hamiltonians in the Table we make the following observations:
The classical/traditional Hamiltonians of particle/light optics are
modified by wavelength-dependent contributions in the
quantum/non-traditional prescriptions respectively.  The algebraic
forms of these modifications in each row is very similar.  This should
not come as a big surprise.  The starting equations have one-to-one
algebraic correspondence: Helmholtz $\leftrightarrow$ Klein-Gordon;
Matrix form of Maxwell $\leftrightarrow$ Dirac equation.  Lastly, the
de Broglie wavelength, ${\LAMBDA}_0$, and $\LAMBDA$ have an
analogous status, and the classical/traditional limit is obtained by
taking ${\LAMBDA}_0 \longrightarrow 0$ and
$\LAMBDA \longrightarrow 0$ respectively.  The parallel of the
analogies between the two systems is sure to provide us with more
insights.

\section{Hamiltonians in Different Prescriptions}
{\small
\noindent
The following are the Hamiltonians, in the different prescriptions
of light beam optics and charged-particle beam optics for magnetic
systems.  $\widehat{H}_{0\,, p}$ are the paraxial Hamiltonians, with
lowest order wavelength-dependent contributions.

\noindent
\begin{tabular*}{6.0in}[t]{@{\extracolsep{\fill}}|ll|}
\hline
& \\
\parbox[t]{2.5in}{
\parbox[t]{2.5in}{
{\large\bf Light Beam Optics}
}}
&
\parbox[t]{3.0in}{
{\large\bf Charged-Particle Beam Optics}
} \\
& \\
\hline
& \\
\parbox[t]{2.5in}{
{\bf Fermat's Principle} \\

$
{\cal H}
=
- \left\{n^2 (\r) - \p_\perp^2 \right\}^{1/2}
$
}

&

\parbox[t]{2.5in}{
{\bf Maupertuis' Principle} \\

$
{\cal H}
=
- \left\{p_0^2 - {\vpi}_\perp^2 \right\}^{1/2} - q A_z
$
} \\
& \\
\hline
& \\
\parbox[t]{2.5in}{

{\bf Non-Traditional Helmholtz} \\

$
\widehat{H}_{0\,, p} = \\
- n (\r)
+ \frac{1}{2 n_0} \hatp_{\perp}^2 \\
- \frac{\i \LAMBDA}{16 n_0^3}
\left[\hatp_\perp^2 , \ddz n (\r) \right]
$

}

&

\parbox[t]{2.5in}{
{\bf Klein-Gordon Formalism} \\

$
\widehat{H}_{0\,, p} = \\
- p_0 - q A_z + \frac{1}{2 p_0} \widehat{\vpi}_\perp^2 \\
+ \frac{\i \hbar}{16 p_0^4}
\left[\widehat{\vpi}_\perp^2 \,, \ddz \widehat{\vpi}_\perp^2 \right]
$
} \\
& \\
\hline
& \\

\parbox[t]{2.5in}{

{\bf Maxwell, Matrix} \\

$
\widehat{H}_{0\,, p} = \\
- n (\r) + \frac{1}{2 n_0} \hatp_\perp^2 \\
- \i \LAMBDA \beta
{\mbox{\boldmath $\Sigma$}} \cdot {\mbox{\boldmath $u$}} \\
+
\frac{1}{2 n_0} \LAMBDA^2 w^2 \beta
$
}

&

\parbox[t]{2.5in}{
{\bf Dirac Formalism} \\

$
\widehat{H}_{0\,, p} = \\
- p_0 - q A_z  + \frac{1}{2 p_0} \widehat{\vpi}_\perp^2 \\
- \frac{\hbar}{2 p_0}
\left\{\mu \gamma
{\mbox{\boldmath $\Sigma$}}_\perp \cdot \B_\perp
+
\left(q + \mu \right) \Sigma_z B_z \right\} \\
+ \i \frac{\hbar}{m_0 c} \epsilon B_z
$
} \\
& \\
\hline
\end{tabular*}

\bigskip

\noindent
{\bf Notation}

\noindent
\begin{tabular*}{6.0in}[t]{@{\extracolsep{\fill}}ll}
\parbox[t]{2.5in}{
\parbox[t]{2.5in}{
$
{\rm Refractive ~ Index}, ~
n (\r) = c \sqrt{\epsilon (\r) \mu (\r)} \\
{\rm Resistance}, ~
h (\r) = \sqrt{{\mu (\r)}/{\epsilon (\r)}} \\
{\mbox{\boldmath $u$}} (\r)
=
- \frac{1}{2 n (\r)} \Nab n (\r) \\
{\mbox{\boldmath $w$}} (\r)
=
\frac{1}{2 h (\r)} \Nab h (\r) \\
$
${\mbox{\boldmath $\Sigma$}}$ and $\beta$ are the Dirac matrices.
}}

&

\parbox[t]{2.5in}{
$
\widehat{\vpi}_\perp
= {\widehat{\mbox{\boldmath $p$}}}_\perp
- q {\mbox{\boldmath $A$}}_\perp \\
\mu_a ~ {\rm anomalous ~ magnetic ~ moment}. \\
\epsilon_a ~ {\rm anomalous ~ electric ~ moment}. \\
\mu = {2 m_0 \mu_a}/{\hbar}\,, ~~~~
\epsilon = {2 m_0 \epsilon_a}/{\hbar} \\
\gamma = {E}/{m_0 c^2}
$
}
\end{tabular*}

}

\end{document}